\documentclass[aps,prl,twocolumn,showpacs,amsmath,amssymb,superscriptaddress]{revtex4-1}
\usepackage{graphicx}
\usepackage{color}
\usepackage{accents}
\usepackage[labeled,resetlabels]{multibib}
\newcites{S}{ }

\usepackage[normalem]{ulem}

\def\be{\begin{equation}}
\def\ee{\end{equation}}
\def\ba{\begin{eqnarray}}
\def\ea{\end{eqnarray}}
\def\bc{\begin{center}}
\def\ec{\end{center}}

\newcommand{\normord}[1]{%
{:\mathrel{\mspace{1mu}#1\mspace{1mu}}:}
}

\begin{document}
\title{Spectra of ultrabroadband squeezed pulses and the finite-time Unruh-Davies effect}

\author{T.L.M. Guedes}

\email{thiago.lucena@uni-konstanz.de}

\affiliation{Department of Physics and Center for Applied
Photonics, University of Konstanz, D-78457 Konstanz, Germany}

\author{M. Kizmann}

\affiliation{Department of Physics and Center for Applied
Photonics, University of Konstanz, D-78457 Konstanz, Germany}

\author{D.V. Seletskiy}

\affiliation{Department of Physics and Center for Applied
Photonics, University of Konstanz, D-78457 Konstanz, Germany}
\affiliation{Department of Engineering Physics, Polytechnique Montré\'{e}al, H3T 1J4, Canada}

\author{A. Leitenstorfer }

\affiliation{Department of Physics and Center for Applied
Photonics, University of Konstanz, D-78457 Konstanz, Germany}

\author{Guido Burkard}

\affiliation{Department of Physics and Center for Applied
Photonics, University of Konstanz, D-78457 Konstanz, Germany}

\author{A.S. Moskalenko}

\email{andrey.moskalenko@uni-konstanz.de}

\affiliation{Department of Physics and Center for Applied
Photonics, University of Konstanz, D-78457 Konstanz, Germany}

\date{\today}

\begin{abstract}
We study spectral properties of quantum radiation of ultimately short duration. In particular, we introduce a continuous multimode squeezing operator for the description of subcycle
 pulses of entangled photons generated by a coherent-field driving in a thin nonlinear crystal with second order susceptibility. We find the ultrabroadband spectra of the emitted quantum radiation perturbatively in the strength of the driving field. These spectra can be related to the spectra expected in an Unruh-Davies experiment with a finite time of acceleration. In the time domain, we describe the corresponding behavior of the {normally ordered} electric field variance.
\end{abstract}


\maketitle

\textit{Introduction.}---In quantum optics, parametric down-conversion (PDC) in nonlinear crystals (NXs) has been {routinely} used to generate pairs of monochromatic entangled {photons~\cite{Wu1986,Kwiat}.} The so obtained squeezed states of light have found applications in a broad range of areas like gravitational wave detection~\cite{interferometer, enhanced_grav_det}, quantum communication systems~\cite{QKD_squeeze, quan_tech, Gisin} and precision measurements~\cite{xiao1987precision,quantum_measurements}. The active interest in squeezed states can be mainly related to the fact that the variance of a given phase space quadrature (a quantum-optical analogue of a canonical variable) is lower for a squeezed state than for a coherent state, including the vacuum state itself. In order to fulfill Heisenberg's uncertainty principle the variance of the conjugate quadrature exhibits the opposite behavior.

In recent years, theoretical and experimental efforts have been made to describe and generate multimode squeezed states~\cite{Squeezing_Wasilewski,blow1990continuum, Yaakov, Silberhorn_PDC_FC, Manko, Silberhorn_timemodes}. Although they have already been experimentally realized by a number of groups~\cite{ Yaakov, Silberhorn_timemodes}, most of the achievements so far are limited to squeezed states with relatively narrow spectra, where the central frequency approximation is still valid. New developments in ultra-stable few-cycle laser sources and advanced detection techniques have paved the way for the generation of few-cycle
pulses of mid-infrared (MIR) squeezed light and the electro-optic detection of their electric field statistics with subcycle temporal resolution~\cite{vacuum_samp,Moskalenko,subcycle}.
The subcycle features in the noise patterns of the generated quantum fields are due to the spatio-temporal modulation of the refractive index of the NX induced by the driving field~\cite{subcycle}. This is analogous to a time-dependent metric for the space-time occupied by the electric field, which leads to photon creation in the perspective of a moving observer~\cite{Birrell}.

The spectral properties of ultrabroadband squeezed states are also of particular interest because they can elucidate
connections between quantum gravitational effects and their table-top optical analogues.
%As an example one might cite
A characteristic example is the Unruh-Davies effect~\cite{Unruh, Davies}, according to which an observer in a non-inertial reference frame, moving with constant acceleration in the vacuum of an inertial reference frame, should detect thermal radiation. This  phenomenon is closely related to the Hawking radiation believed to be emitted at the horizon of black holes~\cite{Hawking}.

The direct observation of these predictions is at the present time infeasible due to technological limitations, and thus optical counterparts were {proposed~\cite{Yablonovitch,Leonhardt,Belgiorno2010,Belgiorno2011,Linder2016,china}} as a means of studying the physics behind such effects.
%There has been,
However, little attention has been paid to the effects of the
%, few
%interest in the effects of the
unavoidably finite (and often short) duration of the effective acceleration experienced by either the light or the detector in the suggested experiments.
%, which might be very important for the optical counterparts of the relativistic systems of interest.
%make mostly use of monochromatic light and thus miss the broadband character of the studied effect, which should be very important when the actual finite duration of the acceleration is considered.

\begin{figure}
    \centering
    \includegraphics[width=\linewidth]{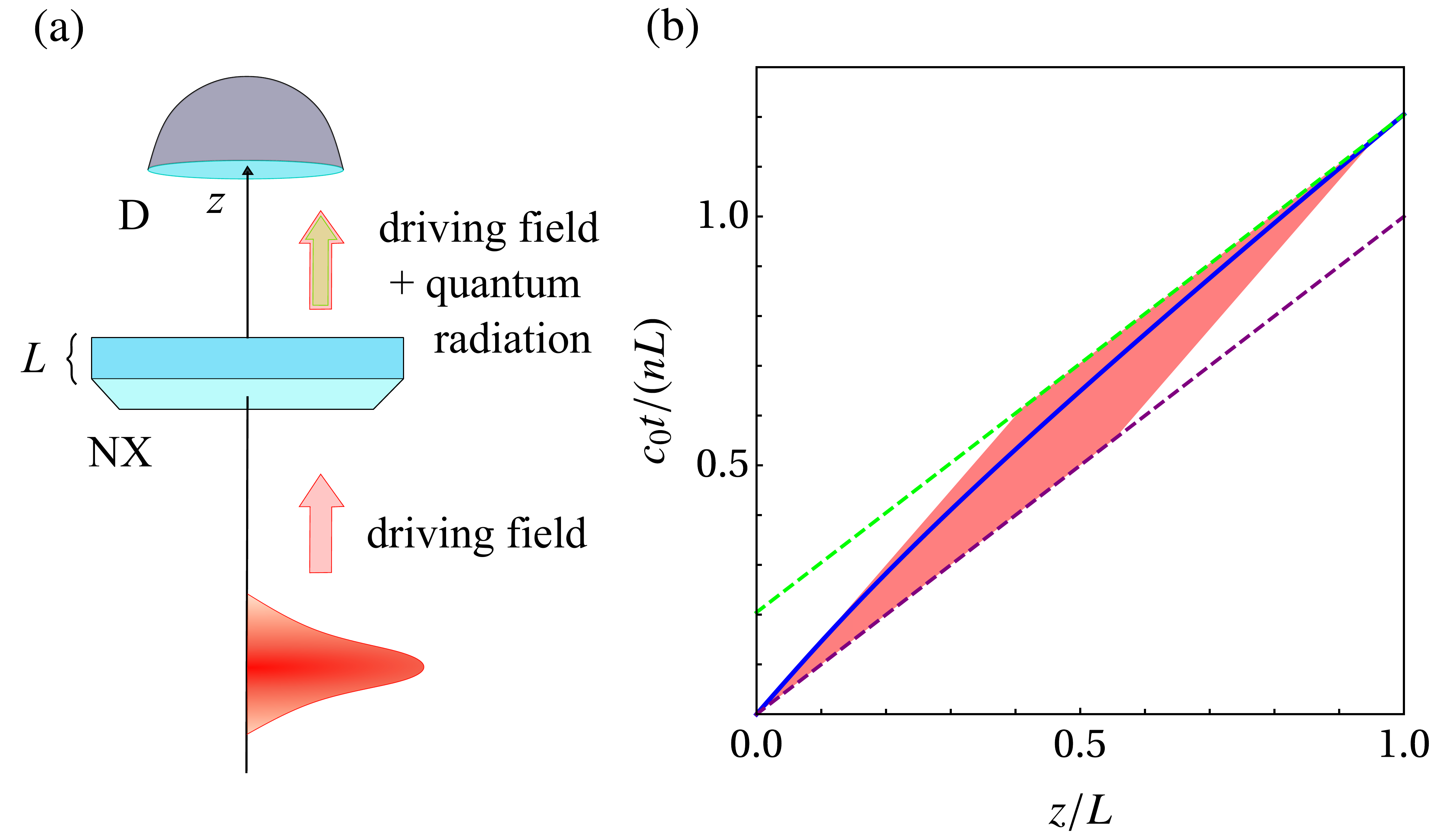}
  \caption{(a) General sketch of the {proposed} experimental setup. The classical driving field propagates through
  %a generating
  a $\chi^{(2)}$ nonlinear crystal (NX) of small thickness $L$ and unperturbed refractive index $n$, generating  ultrabroadband squeezed quantum light. The outgoing light is registered by the detector D. (b) World line of a plane wave mode of the quantum electric field within the NX with refractive index modulated by a half-cycle pulse (HCP). The trajectory (blue) is given by Eq.~\eqref{WL1} with $C_1=0$, $\alpha/n^2=0.49$ and $n\zeta =12$. The dotted purple straight line indicates the trajectory of light in the absence of nonlinear effects. After the acceleration has mostly ceased, the world line approaches the dotted green line parallel to the purple one. The process of acceleration is confined to a diamond-like space-time zone (light red parallelogram) of dimensions defined by the duration of the driving transient.
  }\label{Fig:Worldline}
\end{figure}

In this Letter, we first {introduce} a squeezing operator capable of describing the multimode states generated in a very thin NX with $\chi^{(2)}$ nonlinearity when a coherent ultrashort driving pulse is applied. The relevant experimental setup is schematically shown in Fig.~\ref{Fig:Worldline}(a). Due to the minute thickness of the crystal, phase matching {can be assumed} perfect. The driving pulse
%copropagating through the crystal take part in
induces a nonlinear mixing cascade, with the PDC acting as a seed for the subsequent frequency conversion processes. The superposition of these forms the structure of the emitted quantum field.
%This can be seen from
Next, we study its spectral properties
%of {the resulting quantum states}, which
and confirm the ultrabroadband character of the generated pulses of squeezed light.
%
%of these squeezed pulses.
 Moreover, perturbative calculation of the time-dependent variance of the electric field operator
 %for these states is derived
 %solidifies the connections of %the present
 {links}
our work with related experimental results on subcycle-resolved sampling of the electric field statistics of quantum-optical states~\cite{vacuum_samp,Moskalenko,subcycle}. %From these two results, a method to evaluate the spectrum from the measured time-resolved electric field variance is proposed in the weak field amplitude regime.
 Finally, we make a comparison of the obtained spectra for ultrabroadband squeezed pulses and thermal radiation, aiming to elucidate connections with the Unruh-Davies radiation. It turns out that the limited lifetime of the refractive index perturbation in the crystal %(modulated by the classical electric field that drives the squeezing of the vacuum)
results in spectra with exponentially decaying high-frequency tails, which depend on the duration of the perturbation. This result can be related to the diamond temperature~\cite{diamond_temp, Ralph_diamond} derived for the Unruh-Davies effect when the observer follows an accelerated trajectory during a finite time interval. In our treatment, however, it is the observed incoming vacuum state that undergoes an effective acceleration within a certain space-time zone. This is illustrated in Fig.~\ref{Fig:Worldline}(b) for a plane wave mode of a quantum field that simultaneously enters the crystal with the peak of the driving field.

\textit{Squeezing operator.}---Considering
%We start by deriving the {multimode} squeezing operator to be used in the following calculations.
 previously proposed expressions for a multimode squeezing operator~\cite{blow1990continuum, generalized_sq} and {using} the convention
%possibility of conversion between creation and annihilation operators when the whole continuous frequency domain is taken into account
$\hat{a}(-\omega)=\hat{a}^\dagger(\omega)$ ($\omega\in\mathbb{R}$) connecting creation and annihilation operators for positive and negative frequencies~\cite{MIT}, we introduce the following ansatz for the form of the (continuous) multimode squeezing operator:
\begin{equation}
\hat{S}[\xi]=\exp\bigg\{\frac{1}{2}\Big[\xi^*_{\omega_1,\omega_2}\hat{a}(\omega_1)\hat{a}(\omega_2)
-\xi_{\omega_1,\omega_2}\hat{a}^{\dagger}(\omega_1)\hat{a}^{\dagger}(\omega_2)\Big]\!\bigg\}.
\label{ansatz}
\end{equation}
Here, we employed a generalized Einstein's convention meaning that product terms are integrated from $-\infty$ to $\infty$ over all continuous variables with reoccurring integer indices. For the unitarity of the squeezing operator the frequency-dependent squeezing parameter $\xi_{\omega,\omega'}$ must satisfy $\xi_{\omega,\omega'}=\xi_{\omega',\omega}$, since then $\hat{S}^\dagger[\xi]=\hat{S}[-\xi]$ and hence $\hat{S}[\xi]^\dagger\hat{S}[\xi]=1$.
Rewriting Eq.~\eqref{ansatz} solely in terms of positive frequencies would lead to
four terms in the integrand of the exponent. Two of them correspond to parametric down-conversion (PDC), while the remaining two correspond to frequency-conversion.

In order to calculate expectation values of operators for the states generated by~\eqref{ansatz}, let us {investigate} how $\hat{a}$ and $\hat{a}^\dagger$ transform under $\hat{S}$.
We utilize a common procedure in quantum optics~\cite{Vogel_book} {by introducing an auxiliary operator $\hat{G}[\tilde{z};\xi]=\hat{S}^{\tilde{z}}[\xi]$ for $\tilde{z} \in [0,1]$, which commutes with $\hat{S}[\xi]$. We then define $\hat{a}(\tilde{z};\omega)=\hat{G}^\dagger[\tilde{z};\xi]\hat{a}(\omega)\hat{G}[\tilde{z};\xi]$}
so that $\hat{a}(0;\omega)=\hat{a}(\omega) \,\,\, \text{and} \,\,\, \hat{a}(1;\omega)=\hat{a}'(\omega)=\hat{S}^\dagger[\xi]\hat{a}(\omega)\hat{S}[\xi]$.
The commutator can be calculated using $[\hat{a}(\omega),\hat{a}(\omega ')]=\delta (\omega+\omega ')[\text{sign}(\omega)-\text{sign}(\omega')]/2$ for any $\omega,\omega'\in\mathbb{R}$.
 Differentiating
$\hat{a}(\tilde{z};\omega)$ with respect to $\tilde{z}$ and inserting the expression for $\hat{G}[\tilde{z};\xi]$ leads to
\begin{eqnarray}
&\frac{\partial \hat{a}(\tilde{z};\omega)}{\partial \tilde{z}}
=\Xi_{\omega,\omega_1}\hat{a}(\tilde{z};\omega_1), \label{diff_eq_a}\\
&\Xi_{\omega,\omega'}=-\text{sign}(\omega)\big(\xi_{\omega,-\omega'}-\xi^*_{\omega',-\omega}\big).\nonumber
\end{eqnarray}
This integro-differential operator equation can be solved perturbatively expanding in $\Xi$, resulting in the Bogoliubov transformation
\begin{eqnarray}
&\hat{a}(\tilde{z};\omega)= U_{\omega,\omega_1}(\tilde{z})\hat{a}(\omega_1), \label{Eq:a_transformed_expansion}\\
&U_{\omega,\omega'}(\tilde{z})=\delta(\omega-\omega')+\tilde{z}\Xi_{\omega,\omega'}
+ \frac{\tilde{z}^2}{2!}\Xi_{\omega,\omega_1}\Xi_{\omega_1,\omega'}\nonumber
 + \ldots\;.
\end{eqnarray}
Eq.~\eqref{Eq:a_transformed_expansion} assures the relation $\hat{a}(\tilde{z};-\omega)=\hat{a}^\dagger(\tilde{z};\omega)$.

\textit{Spectra.}---The squeezing process depends on the build-up of electric fields within the NX, which determine the squeezing parameter in Eq.~\eqref{Eq:a_transformed_expansion}. Such interacting fields $\hat{E}(z,t)$ propagating along the $z$-axis in the crystal [see Fig.~\ref{Fig:Worldline}(a)] can be expressed in terms of plane waves confined to a certain transverse area \cite{Moskalenko}, $\hat{E}(z,t)=\hat{E}(z,\omega_1)\exp [-i\omega_1(t-n z/c_0)]$. Here $c_0$ is the speed of light in free space and $n$ is the unperturbed refractive index of the medium.
It can be found~\cite{subcycle} that due to the $\chi^{(2)}$ interaction process  a coherent  mid-infrared (MIR) driving field of sufficiently large amplitude $E_{\text{MIR}}=\langle \hat{E}_{\text{MIR}}\rangle$ with respect to the amplitude of vacuum fluctuations~\cite{vacuum_samp} modulates the quantum contribution $\delta\hat{E}\equiv \hat{E}-E_{\text{MIR}}$ as
\begin{equation}
\frac{\partial \delta\hat{E}(z;\omega)}{\partial z}=\frac{id\omega}{n c_0} E^*_{\text{MIR}}(z;\omega_1-\omega)\delta\hat{E}(z;\omega_1).
\label{field_eq}
\end{equation}
Here $d$ is the effective nonlinear coefficient of the NX, considered to be dispersionless in the relevant frequency range.

We can change now the variable $z\rightarrow \tilde{z}=z/L$ in Eq.~\eqref{field_eq} and use \cite{Loudon_book} $\delta\hat{E}(z;\omega)=i\,\text{sign}(\omega)\sqrt{\frac{\hbar |\omega|}{4\pi \epsilon_0 c_0n A}}\hat{a}(z;\omega)$,
where $A$ is the normalization area, $\hbar$ is the reduced Planck constant and $\epsilon_0$ is the vacuum permittivity. If we consider that $E_{\text{MIR}}(z;\omega)$ does not change appreciably as a function of $z$, comparison of the result with Eq.~\eqref{diff_eq_a}
gives
\begin{equation}
\Xi_{\omega,\omega'}=iC\, \text{sign}(\omega')\sqrt{|\omega\omega'|} E_{\text{MIR}}(\omega-\omega'),
\label{Eq:Xi_result}
\end{equation}
where $C=dL/(n c_0)$. Similar expressions have been used to describe independent frequency-conversion and PDC processes involving light pulses with a well-defined central frequency~\cite{Silberhorn_PDC_FC, Mukamel}.
 Furthermore, the considered phase matching conditions lead to spatially distinguished signal and idler pulses. In contrast, Eqs.~\eqref{ansatz} and~\eqref{Eq:Xi_result} do not rely on the assumption of a bandwidth much smaller than the central frequency. There is also no separation in the propagation direction of the outgoing photons.

\begin{figure}
  \includegraphics[width=0.85\linewidth]{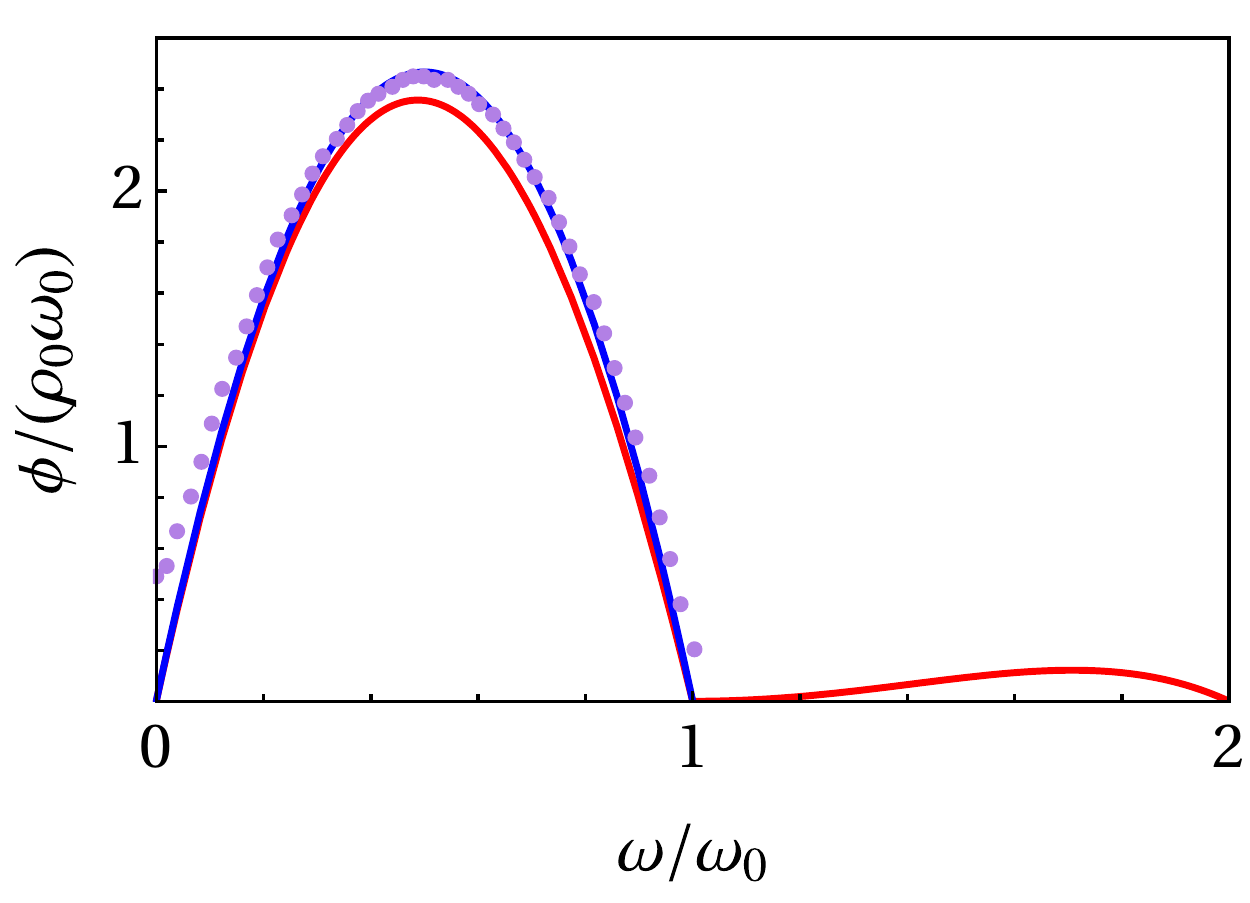}\\
  \caption{{Normalized spectral photon flux density in the case of CW driving ($\rho_0 \omega_0=C^2|E_0|^2\omega^2_0/\pi^2$)}.
  Calculations including up to second (blue) and fourth order (red) terms in $\alpha=dE_0$ have been included.
  The value of the factor $\pi^4\rho_0\omega_0/4$ governing the smallness of the $\alpha^4$ term with respect to the $\alpha^2$ term is 0.02.
  Although the second order contribution has a parabolic shape in the range of $\omega/\omega_0$ from 0 to 1, higher order terms allow for modification of this shape. They also lead to the appearance of photons with frequencies larger than $\omega_0$ and its harmonics.
  The dotted curve shows the average spectral photon flux density for a
%measured during
measurement over a finite time interval $\Delta t = N T$ with $N=50$, where $T=2\pi/\omega_0$ is the period of the driving field.
  }\label{Fig:CW}
\end{figure}

{Using Eqs.~\eqref{Eq:a_transformed_expansion} and~\eqref{Eq:Xi_result} we calculate
{perturbatively}
in $d$ the expectation value of the spectral photon density (SPD) operator, $\hat{\rho}(\omega)=\hat{a}^\dagger(\omega)\hat{a}(\omega)$, for the state $|\{\xi\}_\omega\rangle=\hat{S}[\xi]|0\rangle$ resulting from the {pulse-induced} squeezing process~\cite{Suppl_Mat}.
It is instructive, however, to begin by analysing continuous wave (CW) driving with frequency $\omega_0$, $E_\text{MIR}(\tau)=E_0e^{-i\omega_0\tau}+E^*_0e^{i\omega_0\tau}$}.
Due to the infinite duration of the CW field, the SPD diverges for any frequency of interest.  In this case, {the spectral photon flux density, $\phi(\omega)$, can be defined for a time interval $\Delta t$ and calculated~\cite{Suppl_Mat}\nocite{Thorne, analogue}, as is shown in Fig.~\ref{Fig:CW}}.
We see that PDC is maximally probable near the degeneracy point ($\omega\approx\omega_0/2$), while output at the drive frequency $\omega_0$ is absent.
Additionally, higher order contributions show that the photons generated by PDC can be upconverted to $3\omega_0/2$ by mixing with the
{coherent pump}
field.

\begin{figure}
  \includegraphics[width=0.85\linewidth]{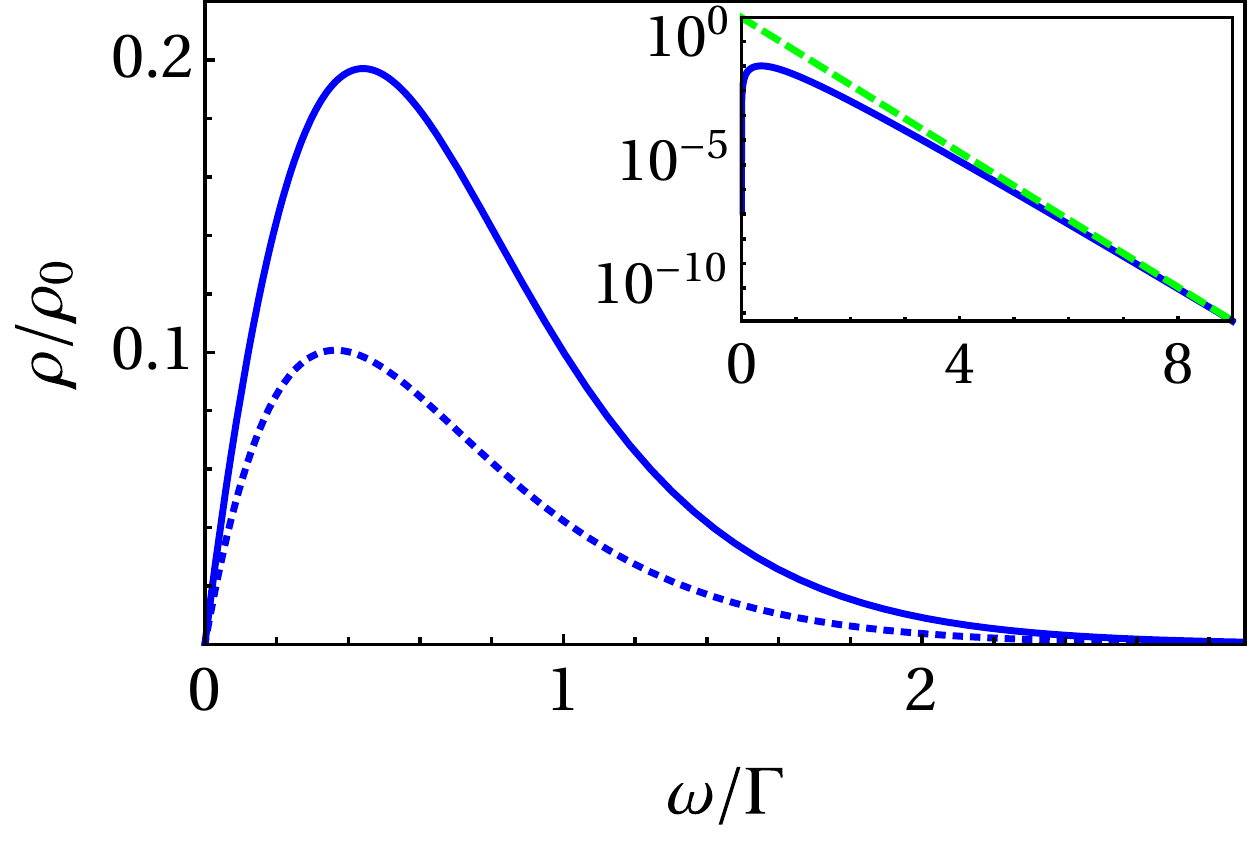}\\
  \caption{{Normalized spectral photon density (SPD) for the driving HCP (dotted blue) and SCP (solid blue) cases ($\rho_0=C^2E^2_0\Gamma/\pi^2$) {in the leading ($\alpha^2$) order.}
%The angular frequencies have been normalized by $\Gamma$, while the photon density distributions were normalized by $\rho_0=C^2E^2_0\Gamma/\pi^2$.
The exponential behavior of the spectra
 %of the {HCP-induced} spectral photon density
 can be better analysed in a logarithmic plot}, presented for the HCP case in the inset (which has the same high-frequency behavior as for the SCP case).
 %on the upper right corner.
 The SPD is shown in blue, while the asymptotic dotted straight line represents a fit of the form $Ae^{-\pi(\omega/\Gamma)}$.
  }\label{Fig:HC_SC}
\end{figure}

Next, we study two cases of the pulsed driving field. Firstly, let us consider an ideal half-cycle pulse (HCP) of light with temporal profile $E_{\text{MIR}}(\tau)=E_0\text{sech}(\Gamma \tau)$ and Fourier transform $E_{\text{MIR}}(\omega)=\frac{E_0}{2\Gamma}\text{sech}(\frac{\pi\omega}{2\Gamma})$~\cite{Moskalenko2}. In this case the SPD
{, $\rho(\omega)=\langle\{\xi\}_\omega|\hat{\rho}(\omega)|\{\xi\}_\omega\rangle$,} reads
%given by:
\begin{equation}
\rho(\omega)=
\frac{C^2E^2_0}{\pi^2} \omega\,\text{ln}\Big(1+e^{-\frac{\pi\omega}{\Gamma}}\Big).
\label{number_sech}
\end{equation}
%Analogously,
For an ideal single-cycle pulse (SCP) of form $E_{\text{MIR}}(\tau)=-E_0(\Gamma \tau)\text{sech}(\Gamma \tau)$, which corresponds to $E_{\text{MIR}}(\omega)=\frac{\pi E_0}{4i\Gamma}\text{sech}(\frac{\pi\omega}{2\Gamma})\text{tanh}(\frac{\pi\omega}{2\Gamma})$ in the frequency domain, we {find}
\begin{equation}
\rho(\omega)=
 \frac{C^2E^2_0}{12}\omega\bigg[\text{ln}\Big(1+e^{-\frac{\pi\omega}{\Gamma}}\Big)+\frac{1}{2}\text{sech}^2\Big(\frac{\pi\omega}{2\Gamma}\Big)\bigg].
\label{number_singcycle}
\end{equation}
%It can be seen from expressions
Both Eqs.~\eqref{number_sech} and~\eqref{number_singcycle} show that for high frequencies the SPD {falls off} as $\exp(-\pi\omega/\Gamma)$,
%both half- and single-cycle pulses up to second order in perturbation theory the spectral photon densities fall as $\exp(-\pi\omega/\Gamma)$ for large frequencies.
i.e., its exponential decay is determined by the duration of the driving field $\Gamma^{-1}$ (see Fig.~\ref{Fig:HC_SC}).
%This behavior will be further addressed later in this letter in connection to the Unruh-Davies radiation.

\begin{figure}
  \includegraphics[width=0.85\linewidth]{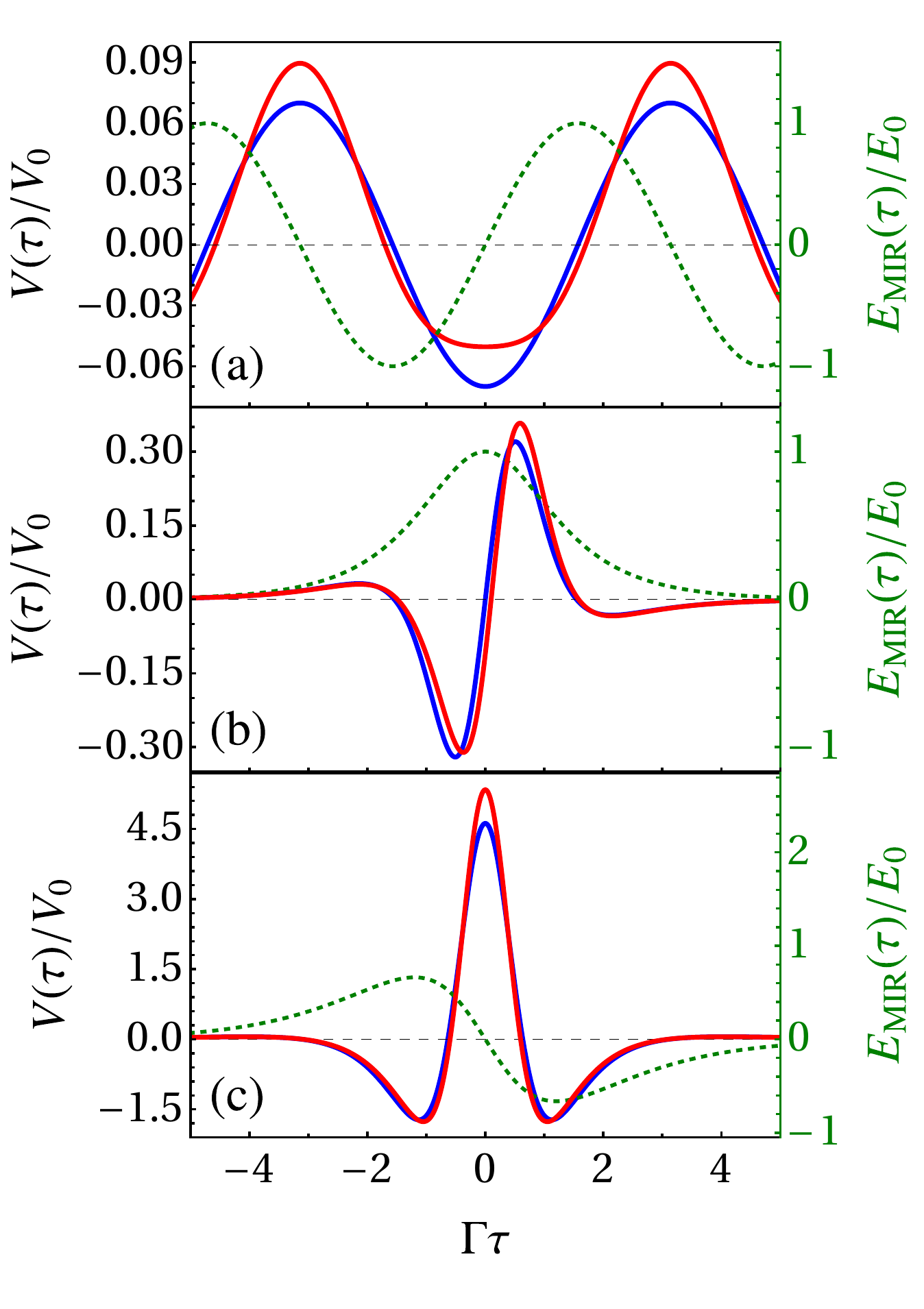}\\
  \caption{{Dynamics} of the normally ordered variance (NOV), $V(\tau)$, of the emitted quantum electric field for (a) CW, (b) HCP and (c) SCP driving {(dotted green).} Contributions up to the first $V^{(1)}(\tau)$ (blue)  and the second $V^{(1)}(\tau)+V^{(2)}(\tau)$ (red) order in the squeezing strength $r$ are shown.
  %=\frac{dL|E_0|\Gamma}{n c_0}$ are shown.
  The %electric field
	NOV is normalized by $V_0=\hbar\Gamma^2/(24\pi\epsilon_0 c_0n A)$, while time is normalized by $\Gamma$ ($\Gamma=\omega_0$ for CW driving). $r=0.07$ for (a), $0.21$ for (b) and $1.54$ {for (c).}
%The shapes of the respective driving transients are shown in the dotted lines (cyan).
}
  \label{Fig:Variance}
\end{figure}

\textit{Electric field variance}.---Another insight into the generated quantum field is provided by its normally ordered variance {(NOV), $V(\tau)$,}
%With the help of Eq.~\eqref{Eq:a_transformed_expansion} we arrive at the result:
which  can be calculated as
$V(\tau)=\langle \{\xi\}_\omega|\normord{[\delta\hat{E}(\tau)]^2}|\{\xi\}_\omega\rangle$
since $\langle \{\xi\}_\omega|\normord{\delta\hat{E}(\tau)}|\{\xi\}_\omega\rangle=0$. Here $\normord{\hat{O}}$ denotes normal ordering for an operator $\hat{O}$~\cite{Vogel_book}.
%Using Eq.~\eqref{prop},
{For the} first order term in the squeezing strength $r=|\alpha|\zeta/n$ %$V^{(1)}(\tau)
%$r=\frac{dL|E_0|\Gamma}{n c_0}$ %$V^{(1)}(\tau)$,
($\alpha=dE_0$, $\zeta=\Gamma L/c_0$)
we obtain~\cite{Suppl_Mat}
\begin{equation}\label{Eq:var1_general}
%\langle \{\xi\}_\omega|\normord{[\Delta \hat{E}(t)]^2}|\{\xi\}_\omega\rangle^{(1)}
V^{(1)}(\tau)=\frac{\hbar C}{24\pi \epsilon_0 c_0 n A}\frac{\partial^3E_{\text{MIR}}(\tau)}{\partial \tau^3}\;.
\end{equation}
The corresponding second order term, $V^{(2)}(\tau)$, was also calculated (for details, see Ref.~\cite{Suppl_Mat}).
%\begin{equation}
%%\begin{aligned}
%\langle \{\xi\}_\omega|\normord{[\Delta \hat{E}(t)]^2}|\{\xi\}_\omega\rangle ^{(2)} =
%   \frac{\hbar C^2 }{2\pi\epsilon c_0 A}\sum_{i=1,2,3}V^{(2)}_i(\tau),
%\end{equation}
%with $V^{(2)}_i$ being three triple integrals involving convolutions of $E_{\text{MIR}}$ with itself (the explicit expressions are given in the supplementary material~\cite{Suppl_Mat_long}). For sufficiently weak driving field amplitudes the first perturbative contribution \eqref{Eq:var1_general} should be enough to describe the behavior of the electric field variance.
The resulting temporal traces for the three aforementioned shapes of $E_\text{MIR}$ are shown in Fig.~\ref{Fig:Variance} for driving field strengths
%effective squeezing strength $\gamma=\frac{dL|E_0|\Gamma}{n c_0}$
such that the perturbation approach is still valid but the impact of the second order contribution becomes visible.

%The variance profiles calculated at the end of the crystal for the three previously studied MIR driving fields are shown in Fig.~\ref{Fig:Variance} for some given values of the squeezing strength $\gamma=\frac{dL|E_0|\Gamma}{n c_0}$.
%The electric field variance for the CW driving case has a periodic profile of alternate electric field squeezing and antisqueezing as a function of time. To lowest order in $\gamma$, the variance profile is an ideal sinusoidal curve, while the next order contribution distorts its shape to increase antisqueezing at the cost of squeezing.
%For the half-cycle driving case the variance profile has a main antisqueezing interval followed by a main squeezing interval. To lowest order in $\gamma$, the variance is antisymmetric relative to the zero measurement time. Higher order contributions break this symmetry in favor of stronger antisqueezing than squeezing.
%The variance for the single-cycle driving case has a unique shape: it is the only studied case in which the maximum squeezing surpasses the antisqueezing at any measured time. Higher order contributions decrease the squeezing while broadening the antisqueezing.
%These results are in good agreement with the temporal variance profiles obtained by \textit{Kizmann et al.} (Kizmann) in the weak squeezing limit.
The dynamics of the NOV is accessible via quantum electro-optic sampling \cite{vacuum_samp,Moskalenko,subcycle} when the time resolution and sensitivity are high enough \cite{Kizmann2018}. Within the range of validity of our perturbation theory, both the NOV and the SPD  are interrelated via the {shape of  $E_\text{MIR}$.}
% [cf. Eqs.~\eqref{prop} and \eqref{Eq:var1_general}].
%
%Expression \eqref{Eq:var1_general} suggests that within the limitations of perturbation theory $E_\text{MIR}$ can be derived from the measured {NOV} of the outgoing pulsed quantum field.
This motivates future experiments to retrieve SPD information %estimated from the data obtained in an ideal
from the temporal traces of the detected field variance obtained via %time-resolved
quantum electro-optic sampling.
%measurement.
%\begin{equation}\label{Eq:V2_1}
%\begin{aligned}
% V^{(2)}_1&=\int^\infty_{0}d\omega \int^\infty_{0}d\omega'|\omega\omega'|
%\int^{\infty}_{0}d\omega_1|\omega_1|\times\\
%&\Re\big[E^*_{\text{MIR}}(\omega+\omega_1)E_{\text{MIR}}(\omega_1+\omega')e^{i(\omega-\omega')t}\big],
%\end{aligned}
%\end{equation}
%\begin{equation}\label{Eq:V2_2}
%\begin{aligned}
% V^{(2)}_2&=\frac{1}{2}\int^\infty_{0}d\omega \int^\infty_{0}d\omega'|\omega\omega'|
%\int^{\infty}_{-\infty}d\omega_1\omega_1\times \\
%&\Re\big[E_{\text{MIR}}(\omega+\omega_1)E^*_{\text{MIR}}(\omega_1-\omega')e^{-i(\omega+\omega')t}\big],
%\end{aligned}
%\end{equation}
%\begin{equation}\label{Eq:V2_3}
%\begin{aligned}
% V^{(2)}_3&=-\int^\infty_{0}d\omega \int^\infty_{0}d\omega'|\omega\omega'|
%\int^{\infty}_{0}d\omega_1|\omega_1|\times\\
%&\Re\big[E_{\text{MIR}}(\omega-\omega_1)E_{\text{MIR}}(\omega_1+\omega')e^{-i(\omega+\omega')t}\big].
%\end{aligned}
%\end{equation}

\begin{figure}
    \centering
    \includegraphics[width=\linewidth]{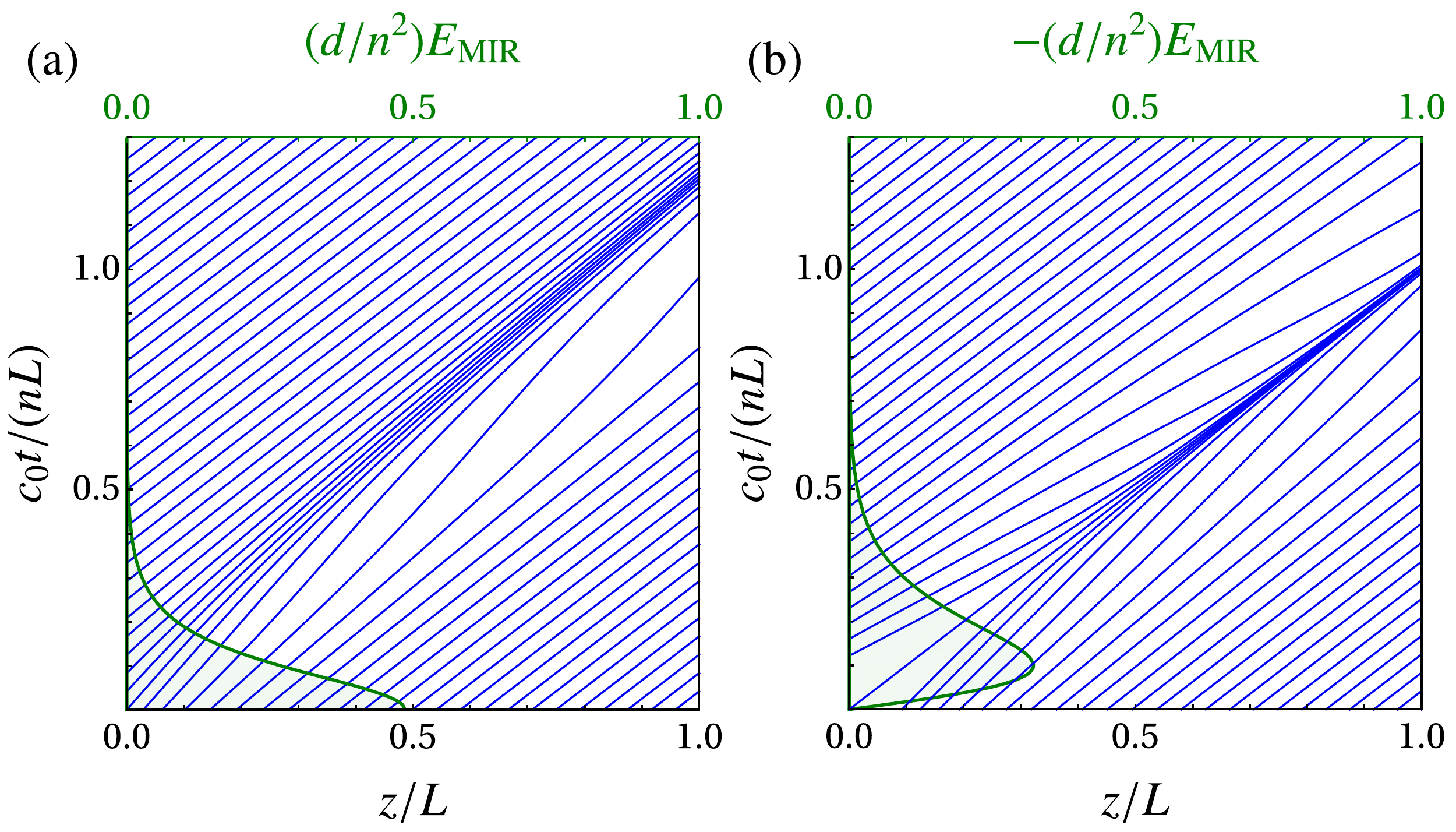}
  \caption{{World} lines of the
   modes of quantum light propagating through the NX
   %vacuum modes' wave fronts
   for HCP (a) and SCP (b) driving. Each world line (blue) is defined by its initial condition, which is given by a certain event at the boundary of the crystal and correspondingly by the amplitude of the driving field {(green)} at this event. Here $\alpha/n^2=0.49$ and $n\zeta=12$
  %.
  %For more details on this derivation
  (see Ref.~\cite{Suppl_Mat}).
  }\label{Fig:Worldlines}
\end{figure}

\textit{Analogue gravity and world lines of light}.---
The quantum properties of the generated light are determined by the effectively curved space-time that the light modes experience while travelling through the NX, dressed by the input driving field. %with spatio-temporal refractive index modulation.
The metrics of such space-time can be extracted from the dispersion relation for the propagating quantum electric field~\cite{Novello, Leo_metrics}.
%
%The {NOVs}
%%electric field variance
%profiles in Fig.~\ref{Fig:Variance}(b) and (c) bear strong resemblances to the density of world lines projected along a line perpendicular to the light rays $t-nz/c_0=\tau = \text{const.}$.
%
This fact allows us to derive the null geodesic equations~\cite{schutz2009} for the %propagating modes of the generated quantum fields
respective modes (for details, see Ref.~\cite{Suppl_Mat}). The world lines follow the equations
\begin{eqnarray}
  \frac{\alpha\zeta}{n}\frac{z}{L}-\text{sinh}\left[\frac{\zeta}{L} (c_0t- n z)\right]&=&C_1, \label{WL1} \\
  \frac{\alpha\zeta}{n} \frac{z}{L}+\text{Chi}\left[\Big|\frac{\zeta}{L} (c_0 t- n z)\Big|\right]&=&C_2 \label{WL2},
\end{eqnarray}
%\begin{equation}
%\frac{\alpha\zeta}{n}\frac{z}{L}-\text{sinh}\left[\frac{\zeta}{L} (c_0t- n z)\right]=C_1,
%\label{WL1}
%\end{equation}
%\begin{equation}
%\frac{\alpha\zeta}{n} \frac{z}{L}+\text{Chi}\left[\Big|\frac{\zeta}{L} (c_0 t- n z)\Big|\right]=C_2,
%\label{WL2}
%\end{equation}
for the HCP and SCP driving cases, respectively. The constants $C_1$ and $C_2$ in these equations define the distance of a propagating wave front relative to the center of the driving field at the entrance of the crystal. $\alpha=dE_0$
%describes
gives the strength of the nonlinear perturbation, $\zeta=\Gamma L/c_0$ determines the spatial extension of the curvature (i.e. acceleration)
relative to the length of the crystal and $\text{Chi}(x)$ is the hyperbolic cosine integral function~\cite{gradshteyn2014}. The world lines for several values of $C_1$ and $C_2$ are shown in Fig.~\ref{Fig:Worldlines} alongside with a projection of the normalized driving fields. The
acceleration of the modes is confined to finite regions of space-time. Moreover, the density of world lines projected along a line perpendicular to the light rays $t-nz/c_0=\tau = \text{const}$ determines the effective change in the flow of time and thereby is connected with the temporal profiles of the detected variance~\cite{Kizmann2018}, as can be comprehended through comparison between Figs.~\ref{Fig:Variance}(a,b) and~\ref{Fig:Worldlines}(a,b). The evolution of the modes in the space-time
curved due to a spatio-temporal varying refractive index leads to
the generation of quantum radiation. It is thus insightful to discuss
 %One can infer that
 %Since the evolution of the modes in the space-time curved due to a varying refractive index leads to the generation of quantum radiation. This motivates us to compare
 our results in relation to one of the most well-known examples of {the creation of quantum light} from the vacuum through acceleration: the Unruh-Davies effect.

\textit{Unruh-Davies radiation and the diamond temperature}.---%Let us compare the SPDs of thermal radiation and ultrabroadband squeezed states.
From Planck's law, the SPD of thermal radiation is dominated by $\exp(-\frac{\hbar \omega}{k_{B} T})$ at large frequencies,  with a decay dependent on the temperature $T$ ($k_{B}$ is Boltzmann's constant).
%for thermal states has an angular frequency dependence proportional to $(\exp(\frac{\hbar \omega}{{k_\mathrm{B}} T})-1)^{-1}$, where ${k_\mathrm{B}}$ is the Boltzmann constant and $T$ is the absolute temperature.
%At large frequencies Planck's law is dominated by $\exp(-\frac{\hbar \omega}{{k_\mathrm{B}} T})$, with a decay dependent on $T$.
According to the Unruh-Davies effect~\cite{Unruh, Davies}, for the quantum radiation detected in the reference frame of a uniformly accelerated observer, moving in the vacuum of an inertial observer (Minkowski vacuum), the temperature is given by $T_{U}=\hbar a/(2\pi k_{B} c_0)$. Here $a$ is the acceleration measured in the accelerated observer's reference frame.

 In the context of an optical analogue of the Unruh-Davies effect, the detector remains at rest while the light follows an accelerated trajectory within a nonlinear material with time-varying refractive index. If one employs a crystal with $\chi^{(2)}$ nonlinearity as such a material, the refractive index can be modulated by
a
coherent driving~\cite{Glauber}
%low frequency
MIR field %. The dependence of the refractive index on the source of modulation is given by the
 through the Pockels effect. To lowest order in $E_\text{MIR}$ the acceleration of the quantum light modes within the crystal depends on its time derivative~\cite{Leonhardt}. From Eqs.~\eqref{number_sech} and \eqref{number_singcycle}, however, it is possible to see that the exponential decay depends only on its duration $\Gamma^{-1}$.

 Martinetti and Rovelli~\cite{diamond_temp} considered the case of an observer with finite lifetime {$\mathcal{T}$} uniformly accelerated in the vacuum of an inertial observer. In this case the Minkowski vacuum is observed as a thermal state with {time-dependent} temperature
 %$T=\hbar L_D a^2/[2\pi k_{B} c^3_0(\sqrt{1+a^2L^2_D/c^4_0}-\text{cosh}(a\tau^0/c_0))]$,
 \begin{equation}\label{Eq:T}
    {T=T_U\varepsilon\big/\big[\sqrt{1\!+\!\varepsilon^2}-\sqrt{1\!+\!\varepsilon^2\tilde{t}^2}\,\big], \ \ \varepsilon=2T_U/T_D,}
 \end{equation}
{where $\tilde{t}=2t/\mathcal{T}\in(-1,1)$ is the normalized lab time and $T_{D}=2\hbar/(\pi k_{B} \mathcal{T})$.
Since the observer's trajectory lies within a space-time diamond determined by {$\mathcal{T}$,} $T_{D}$ is termed the diamond's temperature. The minimal value $T_\mathrm{min}$ of $T$, in the middle of the observer's lifetime, should play the dominant role for the high-frequency tail of the emitted photon spectra.
%We have two clear limit cases determined by the value of parameter $\varepsilon$:
For  sufficiently large lifetime or acceleration, $\varepsilon\gg1$ and $T_\mathrm{min}$ coincides with the Unruh-Davies temperature $T_{U}$. In the opposite situation, $\varepsilon\ll1$ and $T_\mathrm{min}\approx T_{D}$, given directly by the lifetime of the accelerated observer.}
%depends mostly on the lifetime of the accelerated observer.
%considered here
%
%For infinite lifetimes ($L_D\to\infty$) the diamond temperature $T_{D}$ and the Unruh-Davies temperature $T_{U}$ are the same.
%However, {for sufficiently small accelerations or lifetimes} $T_{D}\approx \hbar c_0/(2\pi k_{B} L_D)$ and the detected temperature depends mostly on the lifetime of the accelerated
%observer.
This result was reinforced through
analysis of a two-level detector model with a properly scaled Hamiltonian, which for a finite measurement time reveals that the detected temperature should be  {$T_{D}$~\cite{Ralph_diamond}.}

 %where $L_D$ stands for the distance travelled by light during half of the observer's lifetime and $\tau^0$ is the proper time of the observer measured relative to the middle of his lifetime interval. Since the observer's trajectory lies within a space-time diamond of dimension $L_D$, $T_{D}$ is termed the diamond's temperature.

%In this sense
{The analysis of the present work holds when $\varepsilon$ is small enough~\cite{Suppl_Mat}. The}
spectra of the outgoing quantum light in our analogue optical system should decay as $\exp(-\frac{\hbar \omega}{k_{B} T})$ with a temperature related to the duration of $E_\text{MIR}$, since it dictates the duration of the acceleration of light within the NX. This result is reflected in Eqs.~\eqref{number_sech} and \eqref{number_singcycle} through the decay dependence on $\Gamma$. This can also be seen qualitatively in Fig.~\ref{Fig:Worldline}(b) and Fig.~\ref{Fig:Worldlines}, where curved world lines are confined to certain space-time zones. The same does not happen in the case of a CW driving field, since the respective electric field has {no  defined time duration.}
%neither has a defined time duration nor an average acceleration.

\textit{Conclusions.}---we propose a generalized squeezing operator to describe ultrabroadband squeezed pulses generated in thin $\chi^{(2)}$ NXs by MIR coherent driving fields. We analyse the spectral properties of these squeezed states for three different shapes of the driving field and connect these results to the time-dependent NOV of the electric field operator.
%The results for {the temporal NOV traces} have been compared with theoretical predictions using a different approach and showed good agreement in the weak driving field regime (matthias).
Ultimately, %we compare the obtained spectral photon densities for the ultrabroadband squeezed pulses with the one expected for pure thermal radiation. The results can be related if one accounts for the finite duration of the driving THz pulses, which could be related to the diamond's temperature in an Unruh-Davies-like effect with a finite lifetime for the observer.
we account for the finite duration of the driving MIR pulses and relate
our results to the diamond's temperature in an Unruh-Davies-like effect with a finite lifetime for the observer.

%\vspace{0.2cm}
\begin{acknowledgments}
Funding by the DFG within SFB 767 and the Baden-W\"{u}rttemberg Stiftung via
the Eliteprogramme for Postdocs (project ``Fundamental aspects of relativity and causality in time-resolved quantum optics '') as well as by the  LGFG PhD fellowship program and Young Scholar Fund of the University of Konstanz  is gratefully acknowledged. The authors thank Dr. Takayuki Kurihara, Prof. Dr. Rudolf Haussmann, Philipp Sulzer, Maximilian Russ and Matthew Brooks for the fruitful and elucidating discussions.
\end{acknowledgments}

%\bibliography{report}

%

%\end{document}

\onecolumngrid
\clearpage

\setcounter{equation}{0}
\setcounter{figure}{0}
\setcounter{table}{0}
\setcounter{page}{1}

\makeatletter

\renewcommand{\theequation}{S\arabic{equation}}
\renewcommand{\thefigure}{S\arabic{figure}}
\renewcommand{\bibnumfmt}[1]{[S#1]}
\renewcommand{\citenumfont}[1]{S#1}

\setcounter{secnumdepth}{3}
\renewcommand\thesection{\arabic{section}}

\begin{center}
\textbf{\large \underline{Supplemental Material}}\\[0.5cm]
\textbf{\large Spectra of ultrabroadband squeezed pulses and the finite-time Unruh-Davies effect}
\end{center}

\begin{center}
T.L.M. Guedes, M. Kizmann, D.V. Seletskiy, A. Leitenstorfer, G. Burkard and A.S. Moskalenko
\end{center}

\section{Spectral photon density and spectral photon flux density}

Using the expressions for the transformed creation and annihilation operators defined by Eqs.~\eqref{Eq:a_transformed_expansion} and~\eqref{Eq:Xi_result} we calculate up to second order in $d$ the expectation value of the operator $\hat{a}^\dagger(\omega)\hat{a}(\omega')$ for the states $|\{\xi\}_\omega\rangle=\hat{S}[\xi]|0\rangle$ resulting after the squeezing process,
\begin{equation}\label{prop}
\begin{aligned}
\langle \{\xi\}_\omega|\hat{a}^\dagger(\omega)\hat{a}(\omega')|\{\xi\}_\omega\rangle & = \theta(-\omega)\theta(-\omega')\delta (\omega-\omega')
 -iC\Theta_-(-\omega,-\omega') E_{\text{MIR}}(\omega'\!-\!\omega) \\
&	\hspace{-3.5 cm} -\frac{C^2}{2}
\Theta_+(-\omega,-\omega')\int^{\infty}_{-\infty}\hspace{-0.4cm}d\tilde{\omega}\tilde{\omega} E^*_{\text{MIR}}(\omega\!+\!\tilde{\omega})E_{\text{MIR}}(\tilde{\omega}\!+\!\omega') +C^2\sqrt{|\omega\omega'|}\int^{\infty}_{0}\hspace{-0.4cm}d\tilde{\omega}\tilde{\omega}E^*_{\text{MIR}}(\omega\!+\!\tilde{\omega})E_{\text{MIR}}(\tilde{\omega}\!+\!\omega'),
\end{aligned}
\end{equation}
where $\Theta_{\pm} (\omega,\omega')=\sqrt{|\omega\omega'|}\left[\theta(\omega)\text{sign}(\omega)\pm\theta(\omega')\text{sign}(\omega')\right]$. The spectral photon density, $\rho (\omega)$, is defined from Eq.~\eqref{prop} by setting $\omega=\omega'>0$.

In order to circumvent the problem of the diverging spectral photon density in the case of the CW driving, we can use a different set of creation and annihilation {operators~\citeS[pp. 187-191]{Vogel_bookS}} for a finite observation time $\Delta t$. For $\mu \in \mathbb{Z}^+$ we define $\hat{a}_\mu=-i\sqrt{\frac{4\pi\epsilon c_0 A}{\hbar \omega_\mu \Delta t}}\int^{\Delta t}_0 dt \delta\hat{E}^{(+)}(z,t)\exp (i\omega_\mu \tau)$, with $\omega_\mu=2\pi \mu/\Delta t$, $ \delta\hat{E}^{(+)}(z,t)=\int^\infty_0 d\omega \delta\hat{E}(z;\omega)\exp (-i\omega \tau)$ and $ \Delta t=N T$ [$T$ is the period of $E_\text{MIR}(\tau)$ and $\tau=t-nz/c_0$ is the retarded time]. The larger the number $N$ of periods observed, the higher the frequency resolution. The photon number in mode $\mu$ reads $\langle \hat{n}_\mu \rangle = C^2|E_0|^2\frac{\Delta t}{\omega_\mu} \int^{\omega_0}_0 d\omega \omega^2(\omega_0-\omega)\text{sinc}^2[\frac{\pi N}{\omega_0}(\omega_\mu-\omega)]$ for $\omega\in (0,\omega_0)$. Dividing $\langle \hat{n}_\mu \rangle$ by $\Delta t \Delta\omega$ ($\Delta\omega=\omega_0/N$), we obtain the time-averaged spectral photon flux density measured over the time interval $\Delta t$, $\langle \hat{\phi}_\mu \rangle$. In the limit of $N\to \infty$ this expression is exactly the function multiplying the delta function in Eq.~\eqref{prop} for CW driving when $\omega'\to\omega$. %, i.e. $\text{lim}_{\omega'\to\omega} \langle \hat{a}^\dagger(\omega)\hat{a}(\omega') \rangle %= C^2|E_0|^2\omega_0\theta (\omega_0-\omega) \omega(\omega_0-\omega)\delta(\omega-\omega')
%=(\text{lim}_{N\to\infty}\langle \hat{\phi}_\mu \rangle)\Delta t_{\infty}=\langle \hat{\phi}(\omega)\rangle\Delta t_{\infty}$, with $\Delta t_{\infty}$ being the infinite measurement time given by the delta function.
The limiting spectral photon flux density (the average distribution of generated photons in the frequency domain per unit of time and frequency) including terms up to the fourth order in $\alpha=dE_0$ reads:
\begin{equation}
\begin{aligned}
{\phi(\omega)\equiv}\langle\hat{\phi}(\omega)\rangle = &C^2|E_0|^2\omega (\omega_0-\omega)\theta(\omega_0-\omega)
%\\
%& \hspace{-0.8 cm}
+\frac{C^4}{4}|E_0|^4\theta(2\omega_0-\omega)\omega(2\omega_0-\omega)(\omega_0-\omega)^2\\
%& \hspace{-0.8 cm}
&+\frac{C^4}{3}|E_0|^4\theta(\omega_0-\omega)\omega(3\omega^3-6\omega_0\omega^2+5\omega^2_0\omega-2\omega^3_0) .
\label{CW}
\end{aligned}
\end{equation}

%\section{Comparison between second and fourth order terms in
%the perturbative expansion of the spectral photon density}
\section{Convergence of the perturbative expansion, high-frequency tail and role of the pulse shape}

The validity of restricting our calculations of the spectral photon density to second order terms can be verified by analysing the structure of higher order terms in {$\rho(\omega)$} ($\omega > 0$):
\begin{equation}
\begin{aligned}
 {\rho(\omega)}=& C^2\int^\infty_0 d\omega ' \omega \omega ' |E(\omega +\omega ')|^2 - C^3\Re \bigg\{i\int^\infty_{-\infty}d\omega '\int^\infty_0 d\omega ''\text{sign}(\omega ' )|\omega\omega '\omega ''|E^*(\omega +\omega '')E(\omega +\omega ')E^*(\omega '' -\omega ')\bigg\}\\
&+\frac{1}{3}C^4\Re \bigg\{\int^\infty_{-\infty}d\omega 'd\omega ''\int^\infty_0d\omega '''\text{sign}(\omega '\omega '')|\omega\omega '\omega ''\omega '''|E^*(\omega+\omega ''')E(\omega+\omega ')E^*(\omega '-\omega '')E(\omega ''+\omega ''')\bigg\}\\
&+\frac{1}{4}C^4\int^\infty_{-\infty}d\omega 'd\omega ''\int^\infty_0d\omega '''\text{sign}(\omega '\omega '')|\omega\omega '\omega ''\omega '''|E^*(\omega+\omega ')E(\omega '-\omega ''')E(\omega+\omega '')E^*(\omega ''-\omega ''') + \cdots .
\end{aligned}
\label{n_full}
\end{equation}
{Firstly, we have to answer the question of applicability of the perturbative expansion, in general. Let us consider an ultrashort driving pulse of a single-cycle or subsycle shape and duration $\mathcal{T}=1/\Gamma'$. Normalizing in Eq.~\eqref{n_full} frequencies by $\Gamma'$ and Fourier components of the electric field by $E_0/\Gamma'$, from the structure of expansion \eqref{n_full} we see that it holds only if the condition
\begin{equation}\label{Eq:expansion_validity}
   |CE_0|\Gamma' \sim \varepsilon \frac{L}{L_p}\ll 1
\end{equation}
is fulfilled. Here we used the definitions for $C$ and $\varepsilon=T_U/T_D$ from the main text. $L$ is the crystal length and $L_p\sim c_0/\Gamma'$ is the spatial extent of the driving pulse. Notice that in comparison to the condition separating the two regimes in Eq.~\eqref{Eq:T} an additional factor $L/L_p$ appears in Eq.~\eqref{Eq:expansion_validity}. This is caused by the different spatial regions giving rise to detected photons in the situation described by Martinetti and Rovelli and in our analogue case.}

{Now let us consider the regime when Eq.~\eqref{Eq:expansion_validity} is valid and the perturbation expansion is applicable. We want to analyze the  spectral photon density at $\omega\gg \Gamma'$.
 Does the inclusion solely of the lowest order term ($\propto C^2$) of the expansion  capture the correct high-frequency behavior of the spectral photon density in the high-frequency tail? It turns out that the answer depends on the shape of the pulse.
 A clear indication that the lowest order term is not always sufficient for the studied question is provided immediately by the results demonstrated for the case of CW driving, cf. Eq.~\eqref{CW} and Fig.~\ref{Fig:CW}. The contribution of the $C^2$ term is limited in the frequency domain by the driving frequency $\omega_0$. The frequency conversion processes leading to $C^{2n}$ terms ($n>1$) extend the frequency range for the photon observation to $n\omega_0$. Thus however small $C$ might be, in the case of the CW driving, it is necessary to include the terms of sufficiently large order if high frequencies are considered. The resulting values of the density decay rapidly with the frequency increase but these small values are dominated by  the contributions of higher and higher order.}

{The Fourier transforms of the considered pulsed fields $E(\omega)$ are confined to the frequency range $-\Gamma' < \omega< \Gamma'$, declining rapidly outside of this range. Considering the contributions of $C^{2n}$ terms to the high-frequency tail of the spectral photon density $\rho(\omega)$, there are two competing effects which have to be taken into account: (i) the decay behavior of $E(\omega)$ with $\omega$ and (ii) the increase of the maximum value $\omega^\mathrm{max}_n$ of $\omega$ such that the frequencies $\omega_i$ of all fields $E(\omega_i)$ ($i=1,\ldots,n$)  in the corresponding integrands $E(\omega_1)\ldots E(\omega_n)$ do not leave the interval $(-\Gamma',\Gamma')$. One can see that there are always  terms of the order $C^{2n}$ such that $\omega^\mathrm{max}_n=n\Gamma'$. For $\omega\in(0,n\Gamma')$ the resulting contribution from these terms does not  decay rapidly with increase of frequency.
Without loss of generality, the odd order $C^{2n+1}$ terms  can be excluded from the current discussion since their extension in $\omega$ is also limited by $\omega_\mathrm{max}=n\Gamma'$. With respect to $C^{2n}$ terms, the contribution from $C^{2(n+1)}$ terms extends further in frequency by $\Gamma'$ to $\omega_\mathrm{max}=(n+1)\Gamma'$ but scales down with factor $|CE_0|\Gamma'$.}
%
%
%The term proportional to $C^3$ is equal zero for any pulse shape with real Fourier transform and thus will be neglected in the following analysis.
%
%For a given pulse shape $E(\tau )$ with duration defined by $ \Gamma^{-1}$, the integrations over frequencies get their maximal contributions for frequencies $\omega_A$ and $\omega_B$ such that $-\Gamma < \omega_A + \omega_B < \Gamma$ when the integrand is proportional to $E(\omega_A + \omega_B)$. If one writes the complete set of inequalities for the integration domain that actually contributes to the multiple integrals in Eq.~\eqref{n_full}, it is possible to see that the third term on the right-hand side of Eq.~\eqref{n_full} ($\propto C^4/3$) is limited to $\omega < \Gamma$ (and thus for $0<\omega <\Gamma$ is smaller than the first term by a factor of $\sim C^2E^2_0\Gamma^2$), while the fourth term ($\propto C^4/4$) has contributions up to $\omega = 2\Gamma$.
%
%
%Since the latter is the only higher order term (up to $C^4$) that can surpass the contributions of the $C^2$ term for higher frequencies, one can compare the respective terms by shifting the higher order one by $\Gamma$ and evaluating the ratio of the two terms in the range $0<\omega <\Gamma$.

{We want to compare the impact of the effects (i) and (ii) for the high-frequency tail of $\rho(\omega)$.
The decay due to (ii) is roughly independent of the pulse shape and is given by $\sim \exp (-\omega/\gamma)$,
where the inverse decay constant $\gamma$ can be estimated as  $\gamma=\ln[1/(|CE_0|\Gamma')]\,\Gamma'/2$. Notice that in the considered regime the value of the logarithm is positive.
 In contrast, the character of the decay due to (i) is determined by the pulse shape. In the case when $E(\omega)$ decays as $\exp(-\omega/\Gamma')$, as for the pulse shapes considered in the main text, the lowest order $C^2$ term declines as $\exp(-2\omega/\Gamma')$ for $\omega\gg\Gamma'$. The inverse decay rate $\Gamma'/2$ is lower than $\gamma$ if Eq.~\eqref{Eq:expansion_validity}  is  well fulfilled. Then it is sufficient to include only the $C^2$ term for the study of the high frequency tail of the photon density.
 The situation is different, if $E(\omega)$ vanishes super-exponentially with increasing frequency. For example, Gaussian pulses in the time domain have the Gaussian shape also in the frequency domain and decay as $\exp (-\omega^2/\Gamma'^2)$. For them the effect (ii) would always dominate at high enough frequencies. Higher order contributions must be included for an appropriate description. One can still expect a decay close to exponential, with the inverse decay constant $\sim \gamma$. Notice that if $|CE_0|\Gamma'$ is not too small, which is anyway required in order to avoid having too few photons for detection, $\gamma$ does not deviate much from $\Gamma'$ and remains mainly determined by the pulse duration. }

\section{Normally ordered electric field variance}

Let us now introduce a continuous frequency-dependent quadrature operator $\hat{x}(\omega,\phi)$ for the generated quantum electric field
\begin{equation}
\delta\hat{E}(t,z) =\int^\infty_{0}d\omega\,\sqrt{\frac{\hbar \omega}{4\pi\epsilon c_0 n A}}\hat{x}(\omega,\phi),
\label{quad_def}
\end{equation}
where $\hat{x}(\omega,\phi)=\hat{a}(\omega)e^{i\phi(\omega,t,z)} + \hat{a}^\dagger(\omega)e^{-i\phi(\omega,t,z)}$ and $\phi(\omega,t,z)=-\omega(t-n z/c_0)+\pi/2$.
Equation \eqref{quad_def} enables us to connect the normally ordered variance of $\hat{x}(\omega,\phi)$ to the time-dependent normally ordered variance of the electric field operator.
 From equation \eqref{quad_def} we see that the normally ordered variance $\langle \normord{[\delta \hat{E}(\tau)]^2}\rangle$ depends on the values of $\langle \normord{ \hat{x}(\omega)}\rangle$ and $\langle \normord{ \hat{x}(\omega) \hat{x}(\omega')}\rangle$. For the state $|\{\xi\}_\omega\rangle $, $\langle \{\xi\}_\omega|\hat{x}(\omega)|\{\xi\}_\omega\rangle = \langle 0|\hat{S}^\dagger\hat{x}(\omega)\hat{S}|0\rangle $ whereas $\hat{S}^\dagger\hat{x}(\omega)\hat{S}$ is linear in both $\hat{a}(\omega)$ and $\hat{a}^\dagger(\omega)$, resulting in a zero expectation value. The other expectation value reads
\begin{equation}
\begin{aligned}
\langle \{\xi\}_\omega|\normord{\hat{x}(\omega)\hat{x}(\omega')}|\{\xi\}_\omega\rangle &= \langle \{\xi\}_\omega|\hat{a}(\omega)^\dagger\hat{a}(\omega')|\{\xi\}_\omega\rangle  e^{-i(\phi-\phi')} + \langle \{\xi\}_\omega|\hat{a}^\dagger(\omega')\hat{a}(\omega)|\{\xi\}_\omega\rangle e^{i(\phi-\phi')} \\
 & + \langle \{\xi\}_\omega| \hat{a}(\omega)\hat{a}(\omega')|\{\xi\}_\omega\rangle  e^{i(\phi+\phi')} + \langle \{\xi\}_\omega|\hat{a}^\dagger(\omega)\hat{a}^\dagger(\omega')|\{\xi\}_\omega\rangle  e^{-i(\phi+\phi')}.
\end{aligned}
\label{quadrature_var}
\end{equation}

The four terms in \eqref{quadrature_var} can be calculated by transforming the creation and annihilation operators with the squeezing operator as in Eq.~\eqref{Eq:a_transformed_expansion} and taking the vacuum expectation value, Eq.~\eqref{prop}. %We restrict ourselves to calculations up to the second order in $d$ and consider $\omega,\omega'>0$.
The normally ordered variance thus reads
%\begin{equation}
%\begin{aligned}
%\langle \{\xi\}_\omega|\normord{[\Delta \hat{E}(t)]^2}|\{\xi\}_\omega\rangle & =
% \frac{\hbar }{2\pi\epsilon_0 c_0 A}\int^\infty_{0}d\omega d\omega'|\omega\omega'|
%\bigg\{C\Im\big[E_{\text{MIR}}(\omega+\omega')e^{-i(\omega+\omega')t}\big] \\
%	& +C^2 \int^{\infty}_{0}d\omega_1|\omega_1|\Re\big[E^*_{\text{MIR}}(\omega+\omega_1)E_{\text{MIR}}(\omega_1+\omega')e^{i(\omega-\omega')t}\big] \\
%	& +\frac{1}{2}C^2 \int^{\infty}_{-\infty}d\omega_1\omega_1\Re\big[E_{\text{MIR}}(\omega+\omega_1)E^*_{\text{MIR}}(\omega_1-\omega')e^{-i(\omega+\omega')t}\big] \\
% & -C^2 \int^{0}_{-\infty}d\omega_1|\omega_1|\Re\big[E_{\text{MIR}}(\omega+\omega_1)E_{\text{MIR}}(-\omega_1+\omega')e^{-i(\omega+\omega')t}\big]\bigg\}.
%\end{aligned}
%\label{var}
%\end{equation}
\begin{equation}\label{varS}
\langle \{\xi\}_\omega|\normord{[\delta \hat{E}(\tau)]^2}|\{\xi\}_\omega\rangle =
\langle \{\xi\}_\omega|\normord{[\delta \hat{E}(\tau)]^2}|\{\xi\}_\omega\rangle^{(1)}+\langle \{\xi\}_\omega|\normord{[\delta \hat{E}(\tau)]^2}|\{\xi\}_\omega\rangle^{(2)}=V^{(1)}(\tau)+V^{(2)}(\tau),
\end{equation}
where
\begin{equation}\label{Eq:var1_generalS}
\langle \{\xi\}_\omega|\normord{[\delta \hat{E}(\tau)]^2}|\{\xi\}_\omega\rangle^{(1)}
=\frac{\hbar C}{2\pi  \epsilon c_0 A}\int^\infty_{0}d\omega d\omega'|\omega\omega'| \Im\big[E_{\text{MIR}}(\omega+\omega')e^{-i(\omega+\omega')\tau}\big]
=\frac{\hbar C}{24\pi \epsilon c_0n A}\frac{\partial^3E_{\text{MIR}}(\tau)}{\partial \tau^3},
\end{equation}
and
\begin{equation}
%\begin{aligned}
\langle \{\xi\}_\omega|\normord{[\delta \hat{E}(\tau)]^2}|\{\xi\}_\omega\rangle ^{(2)} =
   \frac{\hbar C^2 }{2\pi\epsilon_0 c_0n A}\sum_{i=1,2,3}V^{(2)}_i(\tau),
%   \\
% & \hspace{-3.5 cm} = \frac{\hbar d^2l^2 }{2\pi\epsilon_0 n^2c^3_0 A}\int^\infty_{0}d\omega d\omega'|\omega\omega'|
%\Big\{\int^{\infty}_{0}d\omega_1|\omega_1|\Re\big\{E^*_{\text{MIR}}(\omega+\omega_1)E_{\text{MIR}}(\omega_1+\omega')e^{i(\omega-\omega')t}\big\} \\
%	& \hspace{-3.5 cm}+ \frac{1}{2}\int^{\infty}_{-\infty}d\omega_1\omega_1\Re\big\{E_{\text{MIR}}(\omega+\omega_1)E^*_{\text{MIR}}(\omega_1-\omega')e^{-i(\omega+\omega')t}\big\} \\
 %& \hspace{-3.5 cm}- \int^{\infty}_{0}d\omega_1|\omega_1|\Re\big\{E_{\text{MIR}}(\omega-\omega_1)E_{\text{MIR}}(\omega_1+\omega')e^{-i(\omega+\omega')t}\big\}\Big\}.
%\end{aligned}
\end{equation}
with
\begin{equation}\label{Eq:V2_1}
 V^{(2)}_1=\int^\infty_{0}d\omega \int^\infty_{0}d\omega'|\omega\omega'|
\int^{\infty}_{0}d\omega''|\omega''|\Re\big[E^*_{\text{MIR}}(\omega+\omega'')E_{\text{MIR}}(\omega''+\omega')e^{i(\omega-\omega')\tau}\big],
\end{equation}
\begin{equation}\label{Eq:V2_2}
 V^{(2)}_2=\frac{1}{2}\int^\infty_{0}d\omega \int^\infty_{0}d\omega'|\omega\omega'|
\int^{\infty}_{-\infty}d\omega''\omega''
\Re\big[E_{\text{MIR}}(\omega+\omega'')E^*_{\text{MIR}}(\omega''-\omega')e^{-i(\omega+\omega')\tau}\big],
\end{equation}
\begin{equation}\label{Eq:V2_3}
 V^{(2)}_3=-\int^\infty_{0}d\omega \int^\infty_{0}d\omega'|\omega\omega'|
\int^{\infty}_{0}d\omega''|\omega''|\Re\big[E_{\text{MIR}}(\omega-\omega'')E_{\text{MIR}}(\omega''+\omega')e^{-i(\omega+\omega')\tau}\big].
\end{equation}

Depending on the shape of $E_{\text{MIR}}$, expressions \eqref{Eq:V2_1}, \eqref{Eq:V2_2} and \eqref{Eq:V2_3} can be evaluated either analytically or numerically.

\section{World line of a propagating electric field mode
in a $\chi^{(2)}$ nonlinear crystal}

We start with the wave equation
\begin{equation}
\frac{\partial^2\delta\hat{E}(z,t)}{\partial z^2}-\frac{n^2}{c^2_0}\frac{\partial^2\delta\hat{E}(z,t)}{\partial t^2}=\frac{1}{c^2_0}\frac{\partial^2}{\partial t^2}\big[dE_{\text{MIR}}(z,t)\delta\hat{E}(z,t)\big]
\label{field_time}
\end{equation}
for the propagation of the vacuum electric field operator within a crystal with refractive index modulated by the MIR coherent field ($E_{\text{MIR}}$). Considering that $\langle 0|\delta\hat{E}|0\rangle =0$%\quad \forall\quad t,z\in \mathbb{R}$
, we use an eikonal approximation to write the electric field operator as~\citeS{ThorneS}
\begin{equation}
\delta\hat{E}(z,t)=\hat{A}(z,t)e^{i\phi (z,t)},
\label{A_phi}
\end{equation}
where the phase $\phi$ changes in space and time much faster than $\hat{A}$. Following the eikonal approximation we define the wave vector and angular frequency fields by
\begin{equation}
k(z,t)\equiv\nabla \phi (z,t) \qquad \text{and} \qquad \omega(z,t)\equiv-\partial_t \phi(z,t).
\end{equation}

Inserting Eq.~\eqref{A_phi} into Eq.~\eqref{field_time} and considering that terms that scale as different powers of $k$ (and/or $\omega$) should vanish independently~\citeS{ThorneS}, we get from the term scaling as $k^2$ the following dispersion relation:
\begin{equation}
k^2(z,t)-\bigg[\frac{n^2}{c^2_0}+\frac{dE_{\text{MIR}}(z,t)}{c^2_0}\bigg] \omega^2(z,t)=0.
\label{dispersion}
\end{equation}
Relation \eqref{dispersion} gives an analogue of the classical Hamiltonian, $H\propto\omega (k,z,t)$, from which the Hamilton equations can be defined considering $k$ to be independent of $z$. From Eqs.~\eqref{A_phi} and \eqref{field_time} the term scaling as $k$ gives
\begin{equation}
\frac{d\hat{A}(z,t)}{d\tau^0}=\frac{\partial\hat{A}(z,t)}{\partial t}+\frac{\omega}{k}\frac{\partial\hat{A}(z,t)}{\partial z}=-\bigg(\frac{d\omega^2}{c^2_0 k}\bigg)\frac{\partial E_{\text{MIR}}(z,t)}{\partial t}\hat{A}.
\label{eq_A}
\end{equation}
In expression \eqref{eq_A} $d/d\tau^0$ is the derivative with respect to the proper time ($d/d\tau^0\equiv u^\alpha\partial_\alpha$, with $u^\alpha$ being the 4-velocity, here with two components only). We see that the amplitude operator propagates along a world line with group velocity $v_g%=\partial \omega/\partial k
\approx\omega/k$.

Considering relation \eqref{dispersion} and its similarity to the light-cone equation $g^{\mu\nu}K_\mu K_\nu=0$ for $K_\mu=(\omega/c_0,-k)$, we can derive the metrics \citeS{analogueS}
\begin{equation}
g^{\mu\nu}=\eta^{\mu\nu}+c^{-2}_0 u^\mu u^\nu h(z,t),
\label{metrics}
\end{equation}
where
\begin{equation*}
h(z,t)=-1+n^2+dE_{\text{MIR}}(z,t),
\end{equation*}
$\eta^{\mu\nu}=\text{diag}(1,-1)$ and $u^\mu=\delta^\mu_0c_0$ is the 4-velocity in the momentarily co-moving reference frame. Differentiation of the dispersion relation by $\partial_\lambda=(c^{-1}_0\partial_t,\partial_z)$ reads
\begin{equation}
\partial_\lambda (g^{\mu\nu}K_\mu K_\nu)=g^{\mu\nu}_{,\lambda} K_\mu K_\nu+2g^{\mu\nu}K_{\mu,\lambda} K_\nu=0,
\label{diff_metrics}
\end{equation}
where $ x^{\mu\nu\cdots\zeta}_{,\lambda}\equiv\partial_\lambda x^{\mu\nu\cdots\zeta}$.

Following the definition of the Christoffel symbol~\citeS{schutz2009S}
\begin{equation}
\Gamma^\gamma_{\beta\mu}= \frac{1}{2}g^{\alpha\gamma}(g_{\alpha\beta,\mu}+g_{\alpha\mu,\beta}-g_{\beta\mu,\alpha}),
\label{Christ_def}
\end{equation}
and knowing that the inverse of the metrics is given by
\begin{equation}
g_{\mu\nu}=\eta_{\mu\nu} -c^{-2}_0u_\mu u_\nu\frac{h}{1+h},
\label{inv_metrics}
\end{equation}
we arrive at the result:
\begin{equation}
\Gamma^\gamma_{\beta\mu}= \frac{1}{2c^2_0(1+h)^2}\bigg[-u^\gamma u_{\beta}\partial_\mu h-u^\gamma u_{\mu}\partial_\beta h+u_{\beta}u_\mu\eta^{\gamma\alpha}\partial_\alpha h+hu^\gamma\Big( -u_\beta\partial_\mu h-u_\mu\partial_\beta h+c^{-1}_0 u_\beta u_\mu\partial_0 h \Big)\bigg].
\label{Christ}
\end{equation}
From equations \eqref{Christ_def} and \eqref{Christ} we explicitly see the symmetry $\Gamma^\gamma_{\beta\mu}=\Gamma^\gamma_{\mu\beta}$.

Since $K_{\mu ,\nu}=K_{\nu,\mu}$ ($K_\mu=-\partial_\mu\phi$) and $g^{\mu\nu}K_{\mu,\lambda}=K^\nu_{,\lambda}=K^{,\nu}_\lambda$, the covariant derivative of the metrics reads
\begin{equation}
g^{\mu\nu}_{;\lambda}\equiv g^{\mu\nu}_{,\lambda} +\Gamma^{\mu}_{\alpha\lambda}g^{\alpha\nu}+\Gamma^{\nu}_{\alpha\lambda}g^{\mu\alpha}= 0, %\quad \longrightarrow \quad g^{\mu\nu}_{,\lambda} =-\Big(\Gamma^{\mu}_{\alpha\lambda}g^{\alpha\nu}+\Gamma^{\nu}_{\alpha\lambda}g^{\mu\alpha}\Big).
\end{equation}
from which we find that
\begin{equation}
g^{\mu\nu}_{,\lambda} =-\Big(\Gamma^{\mu}_{\alpha\lambda}g^{\alpha\nu}+\Gamma^{\nu}_{\alpha\lambda}g^{\mu\alpha}\Big).
\label{cov_d}
\end{equation}

With the help of Eq.~\eqref{cov_d} it is easy to show that Eq.~\eqref{diff_metrics} can be rewritten as $K^\mu K^\lambda_{;\mu}=0$, where $K^\lambda_{;\mu}=K^\lambda_{,\mu}+K^\nu\Gamma^\lambda_{\nu\mu}$ is the covariant derivative of $K^\lambda$. Since $K^\mu=g^{\mu\nu}K_\nu=-g^{\mu\nu}\partial_\nu\phi\propto dx^\mu/d\tau^0$ \citeS{Leo_metricsS}, the relations $K^\mu K^\lambda_{;\mu}=0$ give the null geodesic equations
\begin{equation}
\frac{d^2 x^\alpha}{d(\tau^{0})^2}+\Gamma^{\alpha}_{\mu\beta}\frac{d x^\mu}{d\tau^{0}}\frac{d x^\beta}{d\tau^{0}}=0,
\label{geodesic}
\end{equation}
the components of which can be explicitly written as:
\begin{equation}
\frac{d^2 x^0}{d(\tau^{0})^2}+\frac{1}{1+h}\bigg[-\frac{1}{2}\partial_0 h\Big(\frac{d x^0}{d\tau^{0}}\Big)^2-\partial_1 h\frac{d x^0}{d\tau^{0}}\frac{d x^1}{d\tau^{0}}\bigg]=0,
\label{geodesic1}
\end{equation}
\begin{equation}
\frac{d^2 x^1}{d(\tau^{0})^2}-\frac{1}{2(1+h)^2}\partial_1 h\Big(\frac{d x^0}{d\tau^{0}}\Big)^2=0.
\label{geodesic2}
\end{equation}

In the simple case when $h=n^2-1$ (no driving field), equations \eqref{geodesic1} and \eqref{geodesic2} give straight lines of the form $x^1=Ax^0+B$, with $A$ and $B$ defined by the initial conditions (which of course should assure that $A$ will give the correct speed of light within the medium).

Now lets try the case when $h=n^2-1+dE_0\text{sech}\big[\frac{\Gamma}{c_0}(x^0-nx^1)\big]=n^2-1+\alpha\text{sech}[\zeta(x^0_L-n x^1_L)]$ (with $x^\mu_L=x^\mu/L$), which correspond to a half-cycle driving field. The geodesic equations read:
\begin{equation}
\Big(n^2+\alpha\text{sech}[\zeta(x^0_L-n x^1_L)]\Big)\frac{d^2 x^0_L}{d(\tau^{0})^2}+\alpha\zeta\text{sech}[\zeta(x^0_L-n x^1_L)]\text{tanh}[\zeta(x^0_L-n x^1_L)]\bigg[\frac{1}{2}\Big(\frac{d x^0_L}{d\tau^{0}}\Big)^2-n\frac{d x^0_L}{d\tau^{0}}\frac{d x^1_L}{d\tau^{0}}\bigg]=0,
\label{geodesic1HC}
\end{equation}
\begin{equation}
2\Big(n^2+\alpha\text{sech}[\zeta(x^0_L-n x^1_L)]\Big)^2\frac{d^2 x^1_L}{d(\tau^{0})^2}-\alpha\zeta n\text{sech}[\zeta(x^0_L-n x^1_L)]\text{tanh}[\zeta(x^0_L-n x^1_L)]\Big(\frac{d x^0_L}{d\tau^{0}}\Big)^2=0.
\label{geodesic2HC}
\end{equation}

Knowing that the affine parameter $\tau^0$ parametrizes the world line in the $x^0$-$x^1$ space [i.e. (1+1) space-time] and that the trajectory of the light is supposed to be still time-like within the crystal, we introduce the ansatz: $x^1_L(\tau^0)=f(x^0_L(\tau^0))$. This reduces Eqs.~\eqref{geodesic1HC} and \eqref{geodesic2HC} to:
\begin{equation}
\frac{d f}{dx^0_L}-\frac{n}{n^2+\alpha\text{sech}[\zeta(x^0_L-n f)]}=0,
\label{final}
\end{equation}
\begin{equation}
n\frac{d^2 x^0_L}{d(\tau^{0})^2}+\alpha\zeta \text{sech}[\zeta(x^0_L-n f)]\text{tanh}[\zeta(x^0_L-n f)]\frac{d f}{dx^0_L}\Big(\frac{1}{2}-n\frac{d f}{dx^0_L}\Big)\Big(\frac{d x^0_L}{d\tau^{0}}\Big)^2=0.
\end{equation}

The implicit solution of Eq.~\eqref{final} is
\begin{equation}
\alpha x^1_L-\frac{n}{\zeta}\text{sinh}(\zeta (x^0_L- n x^1_L))=C',
\label{WL}
\end{equation}
where $C'$ is a constant. Equation \eqref{WL} gives the approximate world line of the quantum modes within the nonlinear crystal. For the initial condition $x^1_L(0)=0$, $x^0_L=l$, we get $C'=-\frac{n}{\zeta}\text{sinh}(\zeta l)$. From expression \eqref{WL} we see that in the limit $\alpha\to 0$ (no refractive index modulation), the solutions are straight lines of the form $nx^1_L=x^0_L-l$, i.e. a ray of speed $c_0/n$. For $\zeta\to 0$ (constant refractive index modulation), we get $(n+\alpha/n)x^1_l=x^0_L-l$, once again straight lines, but with a different speed, as expected. Eq.~\eqref{WL} can be {rewritten} as $x^0_L(x^1_L)=nx^1_L+\zeta^{-1}\text{sinh}^{-1}(\alpha\zeta x^1_L/n+\text{sinh}(\zeta l))$, which gives time as a function of the space coordinate. This same curve has been attained by treating the evolution of the electric field operator in Eq.~\eqref{field_time} with the method of characteristics within the slow varying amplitude approximation~\citeS{Kizmann2018S}. This means that the characteristic lines leading to the variance profiles of the electric field are also the world lines for the modes of the respective field.

Proceeding in a similar way for $h=n^2-1+dE_0[\Gamma (x^0-nx^1)]\text{sech}\big[\frac{\Gamma}{c_0}(x^0-nx^1)\big]=n^2-1+\alpha\zeta(x^0_L-n x^1_L)\text{sech}[\zeta(x^0_L-n x^1_L)]$, we arrive at the result:
\begin{equation}
\alpha \frac{z}{L}+\frac{n}{\zeta}\text{Chi}\Big[\Big|\frac{\zeta}{L} (c_0 t- n z)\Big|\Big]=C''.
\label{final2}
\end{equation}
Eq.~\eqref{final2} has similar limiting cases as the example above discussed.

%\bibliographystyleS{prsty}
%\bibliographyS{BibSVAA}

\end{document}